# Magnesium-vacancy optical centers in diamond

Emilio Corte[1,2,3§], Greta Andrini[2,1,3§], Elena Nieto Hernández[1,2,3], Vanna Pugliese[1], Ângelo Costa[4], Goele Magchiels[4], Janni Moens[4], Shandirai Malven Tunhuma[4], Renan Villarreal[4], Lino M.C. Pereira[4], André Vantomme[4], João Guilherme Correia[5], Ettore Bernardi[3], Paolo Traina[3], Ivo Pietro Degiovanni[3,2], Ekaterina Moreva[3], Marco Genovese[3,2], Sviatoslav Ditalia Tchernij[2,1,3], Paolo Olivero[1,2,3], Ulrich Wahl[5*], Jacopo Forneris[1,2,3,*]

[1] Physics Department, University of Torino, Torino 10125, Italy
[2] Istituto Nazionale di Fisica Nucleare (INFN), Sezione di Torino, Torino 10125, Italy
[3] Istituto Nazionale di Ricerca Metrologica (INRiM), Torino 10135, Italy
[4] KU Leuven, Quantum Solid State Physics, 3001 Leuven, Belgium
[5] Centro de Ciências e Tecnologias Nucleares, Departamento de Engenharia e Ciências e Engenharias Nucleares, Instituto Superior Técnico, Universidade de Lisboa, 2695-066 Bobadela LRS, Portugal

**Abstract**
We provide the first systematic characterization of the structural and photoluminescence properties of optically active defect centers fabricated upon implantation of 30-100 keV $Mg^+$ ions in artificial diamond. The structural configurations of Mg-related defects were studied by the emission channeling technique for $^{27}Mg$ implantations performed both at room-temperature and 800 °C, which allowed the identification of a major fraction of Mg atoms (~30-42%) in sites which are compatible with the split-vacancy structure of the MgV complex. A smaller fraction of Mg atoms (~13-17%) was found on substitutional sites. The photoluminescence emission was investigated both at the ensemble and individual defect level in a temperature range comprised between 5 K and 300 K, offering a detailed picture of the MgV-related emission properties and revealing the occurrence of previously unreported spectral features. The optical excitability of the MgV center was also studied as a function of the optical excitation wavelength enabling to identify the optimal conditions for photostable and intense emission. The results are discussed in the context of the preliminary experimental data and the theoretical models available in the literature, with appealing perspectives for the utilization of the tunable properties of the MgV center for quantum information processing applications.

\* Corresponding authors: jacopo.forneris@unito.it, uwahl@ctn.tecnico.ulisboa.pt
§ These authors contributed equally to the work.

1. Introduction

Diamond is a promising material platform for photonic quantum technologies, offering single-photon sources for quantum information processing and sensing schemes based on the optical activity of lattice defects. These systems, commonly known as "color centers", can be engineered upon the controlled introduction of impurities in the diamond crystal structure by ion implantation [1-3]. Besides the well-known and widely investigated negatively-charged nitrogen-vacancy center (NV$^-$), offering unique photophysical properties (photo-stability at room temperature, high quantum efficiency, optically addressable spin properties) for quantum sensing and computing applications [4-9], additional single-photon emitting color centers with appealing features have emerged in the last decade, including group-IV impurities [10,11,14,15-18], noble gases [19-21] and other impurity-related defects [22-24].
Particularly, a recent work has shown a preliminary demonstration of the optical activity of Mg-related color centers in diamond. The available experimental data suggested the formation of an optically active defect (magnesium-vacancy center, MgV in the following) upon Mg ion implantation and annealing, denoted by a sharp emission line at 557.4 nm, high Debye-Weller factor and >0.5 Mcps emission intensity [25] and 2-3 ns radiative lifetime. The results fed the deployment of a detailed numerical ab-initio study of the color center's structure [26], offering an intriguing insight in its opto-physical properties, including the prediction of a large and tunable ground state splitting potentially appealing for quantum information processing purposes.
In this work, we report on a systematic investigation of the MgV color center in diamond. The analysis covers both its structural properties and efficiency of formation as evidenced from determining the lattice sites of implanted Mg using the emission channeling technique, and the optical emission properties, studied in photoluminescence (PL) regime at the ensemble and single-photon emitter level, as a function of temperature and excitation wavelength. The present results extend the preliminary findings available in the scientific literature and offer a contextual evidence to support the theoretical model predicted in Ref. [26], both in terms of the structural analysis of the defect formation in Mg-implanted diamond, and on its PL emission features.

2. Structural properties of Mg-related defects in diamond

For lattice location determination of Mg in diamond, we made use of the electron emission channeling (EC) technique from radioactive isotopes [27-30]. The EC method allows probing the sites of radioactive isotopes in single-crystalline samples, and was recently applied to identify the split-vacancy configuration of implanted $^{121}$Sn inside the SnV complex in diamond [31]. The split vacancy configuration can be pictured as the impurity atom in the center of a divacancy, occupying a position which corresponds to the bond center (BC) site in an unperturbed lattice. The radioactive probe atoms are implanted at low fluences and the emitted $\beta^-$ particles are guided by the crystal potential on their way out of the crystal. A two-dimensional position-sensitive detector (PSD) [28,29] is used to measure the angle-dependent emission yield of electrons in the vicinity of major crystallographic directions, providing patterns which are characteristic for the probe atom lattice location in the sample. In the case of Mg, we used the short-lived $^{27}$Mg ($t_{1/2}$=9.45 min), which was produced at CERN's ISOLDE on-line isotope separator facility by means of bombarding Ti targets with 1.4 GeV protons, followed by out-diffusion, resonant laser ionization and mass separation. More experimental details regarding EC experiments with $^{27}$Mg can be found in Refs. 32, 33. The major lattice sites can be identified by fitting the experimentally observed emission yields by linear combinations of theoretical patterns [28,29,34] calculated for specific positions of the emitter atoms in the lattice. For that purpose we used the many-beam approach [27,28] to calculate the expected emission yields for substitutional (*S*) as well as around 250 interstitial sites in the diamond structure, which are obtained by displacing from the *S* position along <111>, <100>, or <110> directions in steps of around 0.04 Å.

The sample used for the EC studies was a <100> oriented single crystal from ElementSix, termed "SC plate CVD", of size 3.0✕3.0✕0.25 mm³, with nitrogen concentration [N] < 1 ppm. Implantations were performed with 30 keV into a 1-mm diameter beam spot, simultaneously with the measurement of $\beta^-$ emission channeling effects by a 30✕30 mm² PSD placed at a distance of 301 mm from the sample, resulting in the angular resolution of ~0.1° (standard deviation). The fluences used to measure single EC patterns were around 3.5-5✕10$^{11}$ cm$^{-2}$, while the total accumulated fluence at the end of the experiment was 1.1✕10$^{13}$ cm$^{-2}$.

The half life of $^{27}$Mg is too short to perform annealing at high temperatures; with regard to thermal treatment of the sample the only feasible option is varying the implantation temperature. The experimental emission patterns around <110>, <211>, <100> and <111> directions during room temperature (RT) implantation are shown in **Fig. 1a-d**, those for 800 °C implantation in **Fig. 2a-d**. As a first qualitative observation we notice that anisotropies of the EC effects measured from $^{27}$Mg are relatively weak in comparison to effects from other elements implanted under similar conditions into diamond, e.g. those of $^{121}$Sn [31]. This indicates that a considerable fraction of emitter atoms occupies lattice sites of relatively low crystal symmetry. In a first approach for the analysis, the experimental patterns were fitted by allowing two fractions ("2-site fits") of emitter atoms, one on ideal substitutional *S* and another on ideal bond-center (BC) sites, plus a flat contribution, which always needs to be considered in the analysis of EC patterns. The resulting best fit patterns for <110>, <211>, <100> are shown in **Fig. 1e-g** and **Fig. 2e-g**. As was outlined previously [31], in the case of the <111> direction, the patterns from *S* and BC sites have qualitatively similar features, which results in the corresponding site fractions of the best fits being unstable. The theoretical patterns included for the <111> direction in **Figs. 1h** and **2h** are hence not for the best fit results of that direction, but when the relative contributions from *S* and BC sites were fixed at the ratios that were derived from the analysis of the other three directions. As can be seen, in all cases the simulated patterns match the experimental results quite well; the <111> measurements are also compatible with the results from the other directions. Compared to 1-site fits using either *S* or BC positions, the 2-site fits with *S* and BC significantly improved the chi square of fit by 19-42%, and by 17–53%, respectively, while for the <111> directions the improvement was much smaller, 1-3% only.

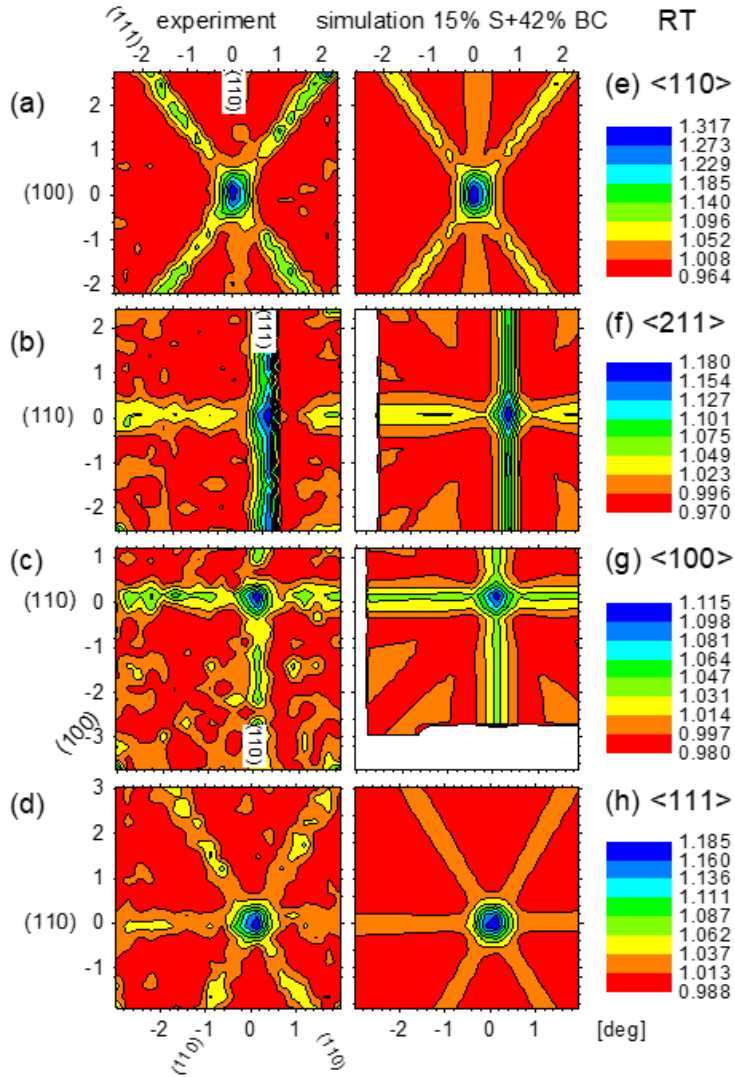

**Figure 1: a)-d)** Experimental $\beta^-$ emission channeling patterns from $^{27}$Mg in diamond around the <110>, <211>, <100>, and <111> directions during RT implantation. The plots **e)-h)** are simulated theoretical patterns considering 15% on ideal substitutional and 42% on ideal bond-center sites. Note that during the RT <211> and <100> measurements the sample was not yet oriented in such a way that the calculated patterns [±3.0°] cover the whole range of measured angles. The areas in the simulated patterns **f)** and **g)** which are not covered, are shown in white.

For both implantation temperatures, the largest fitted fractions were on bond-center sites (42% at RT, 30% at 800 °C), while smaller fractions were assigned to the substitutional positions (15% at RT, 14% at 800 °C). The result implies that 30-40% of implanted Mg atoms are found on sites which are compatible with the theoretically predicted Mg position within the MgV defect in the split-vacancy configuration of $D_{3d}$ symmetry [26]. From a comparison of fluorescence intensity and implanted fluence, the formation efficiency of *optically active* MgV centers had been estimated as 2-3% in undoped diamond implanted at RT and annealed to 800 °C. Our results show that the *structural* efficiency of formation of the split-vacancy configuration in undoped diamond is certainly much higher than a few percent only, suggesting that a large part of the MgV centers are *optically inactive*. The fact that significantly higher optical activation was observed in P-doped diamond, reaching values as high as 48% following 1200 °C annealing [35], indicates that co-doping with the n-type dopant P may transform an inactive form of MgV into optically active MgV centers. Within the scope of the 2-site fit analysis of the EC experimental data, relatively large fractions of emitter atoms (43% at RT, 56% at 800 °C) were assigned to flat contributions to the emission patterns, the so-called "random sites". The assignment of random sites cannot be the consequence of significant radiation damage or amorphization of the sample, since for the light mass of $^{27}$Mg at the applied fluences, the effect of damage should be negligible in diamond, especially for the implantation temperature of 800 °C. The most likely interpretation is that the random fraction represents additional $^{27}$Mg sites of relatively low crystal symmetry, which only produce weak anisotropies in the angular-dependent electron emission yields. Two defect configurations, which could be responsible for such Mg lattice sites, are Mg inside a triple vacancy or inside a quadruple vacancy, which corresponds to the so-called $MgV_2$ or $MgV_3$ complexes. $MgV_2$ was theoretically considered [26], and it was found to have a somewhat higher energy of formation than MgV, which was predicted to be the thermodynamically most stable Mg defect, followed by substitutional Mg. We note, however, that in the case of ion implantation the energy for vacancy creation in the sample is provided by the implantation process, hence the fact that MgV exhibits the lowest defect formation energy may not be the single determining factor which regulates the abundance of complexes formed, and also Mg centers requiring triple or quadruple vacancies might be commonly found. The $MgV_2$ center was predicted to have a similar configuration as MgV, with an additional vacancy "attached from the side" and a symmetry lowering to $C_1$ (although no detailed Mg coordinates were published) [26]. From simple geometrical arguments, one might expect the structure of $MgV_2$ as a central vacancy with two additional nearest-neighbor single vacancies within a (110) plane, and hence with the Mg atom displaced along a <100> direction from an $S$ site. In the case of $MgV_3$, there exist no predictions for its structure or formation energy in the literature. From simple geometric arguments, for Mg inside a quadruple vacancy consisting of one missing C atom with three additional nearest-neighbor single vacancies, one would tentatively expect trigonal symmetry with the Mg atom located close to the so-called anti-bonding (AB) position.

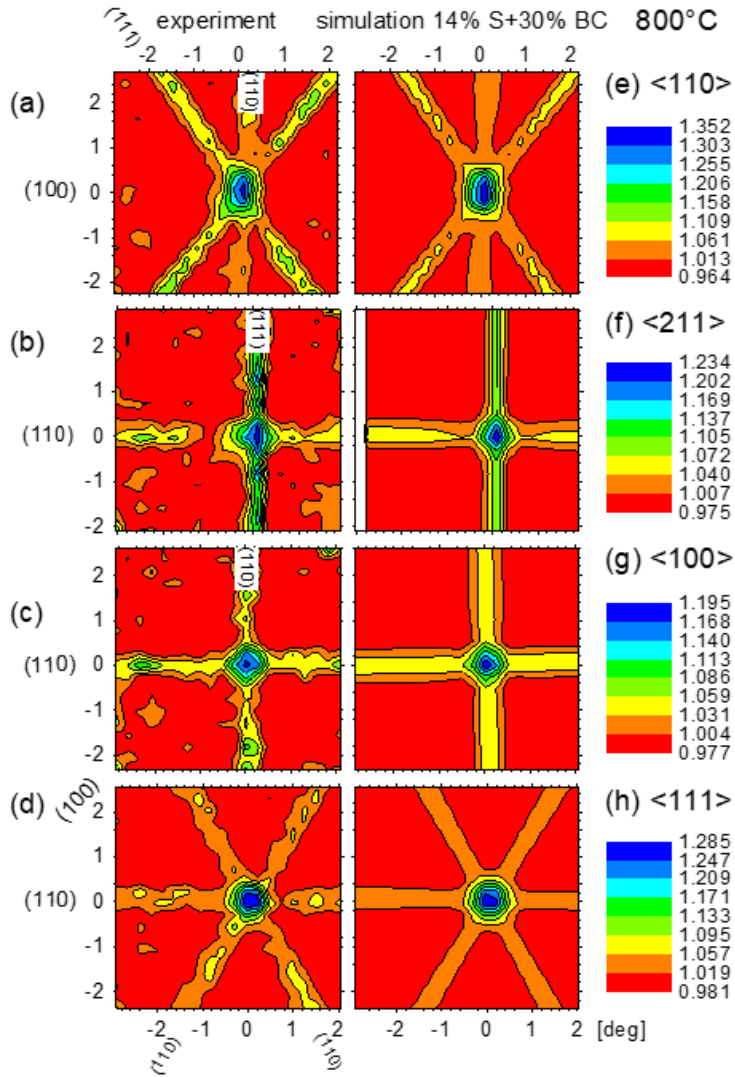

**Figure 2: a)-d)** Experimental $\beta^-$ emission channeling patterns from $^{27}$Mg in diamond around the <110>, <211>, <100>, and <111> directions during 800 °C implantation. The plots **e)-h)** are simulated theoretical patterns considering 14% on ideal substitutional and 30% on ideal bond-center sites.

With these intuitive structures of $MgV_2$ and $MgV_3$ in mind, and also considering the possible existence of isolated interstitial Mg, we performed 3-site fits where, besides Mg on $S$ and BC sites, the position of a third fraction of Mg probes was varied. We did not find any indication for the existence of (isolated) tetrahedral interstitial Mg ($T_d$ symmetry), which supports the theoretical prediction that it has a high formation energy, making it particularly unstable in vacancy-rich ion-implanted diamond. However, while the chi square of <110> and <211> pattern fits improved if Mg sites displaced along <100> from $S$ sites were considered, this result could not be confirmed by the <100> patterns. The identification of a third fraction of Mg probes in highly symmetric lattice sites, in particular displaced from $S$ along <100> directions, was hence not conclusive. An attempt will be made to resolve further Mg lattice sites by means of EC experiments using enhanced angular resolution of 0.05°.

**3. Optical properties of MgV centers**

*3.1 Spectral features of ensemble photoluminescence*

The optical characterization of MgV centers was performed on a set of $2\times2\times0.5$ mm$^3$ IIa single-crystal diamond plates produced by ElementSix by Chemical Vapor Deposition method. The crystals were denoted as "electronic grade" by the supplier, according to the nominal substitutional N and B concentrations below 5 ppb. The sample was implanted with 100 keV $^{24}$Mg$^+$ ions at the IMBL laboratory (KU Leuven). Several squared regions of ~200 µm edge were irradiated at different fluences in the $5\times10^9 - 5\times10^{12}$ cm$^{-2}$ range through the utilization of a custom Al implantation mask. The sample was then processed with a high-temperature thermal annealing (1200 °C, 2 hours at ~$10^{-6}$ hectopascal pressure) and a subsequent oxygen plasma treatment, with the purpose of minimizing the background fluorescence originating from surface contaminants. The room-temperature PL emission spectrum of the highest-fluence Mg-implanted diamond (5x10$^{12}$ cm$^{-2}$) was acquired using a confocal micro-Raman setup under 532 nm excitation wavelength (21.6 mW optical power, **Fig. 3a**). The ensemble emission highlighted the following spectral features:
- a sharp line at 572.8 nm corresponding to the first-order Raman scattering of diamond (1332 cm$^{-1}$ shift);

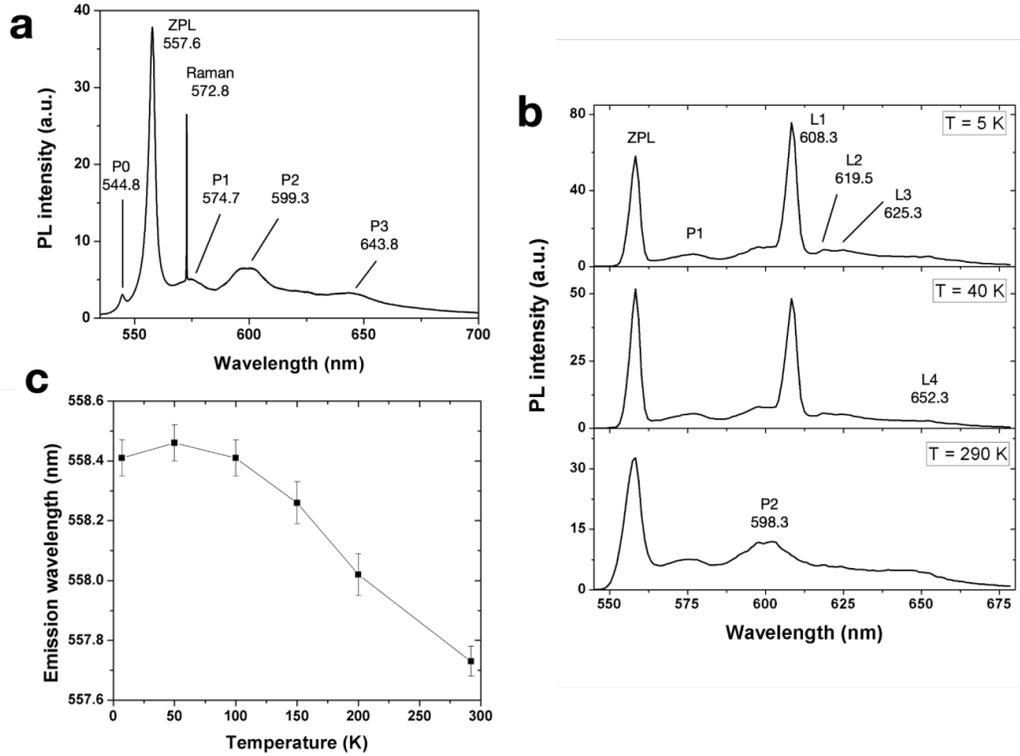

**Figure 3: a)** RT spectrum of MGV ensemble (5x10$^{12}$ cm$^{-2}$ implantation fluence) upon 532 nm laser excitation. **b)** Ensemble spectra acquired under 522 nm excitation (3.7 mW) at 5 K, 40 K, 290 K temperature. **c)** Nominal position of the MgV ZPL dependence on the temperature.

- an intense emission peak at 557.6 nm (2.224 eV; 3 nm FWHM), whose prominence with respect to all the remaining spectral features is in agreement with the early reports on its attribution to the zero-phonon line (ZPL) MgV defect [25] and its theoretical prediction as the main emission of the color center in its negative charge state [26].
- a set of significantly less intense PL peaks centered at 574.7 nm (2.158 eV, denoted as "P1" in the following), 599.3 nm (2.069 eV P2) and 643.8 nm (1.926 eV, P3). The wavelength of these three bands is compatible with the phonon sideband spectrum of the MgV center predicted in Ref [26].
- an additional emission line at 544.8 nm (2.28 eV, "P0" in the following), exhibiting higher emission energy with respect to the MgV ZPL. The origin of such a feature is not clear. Its correlation with Mg-related emission in n-type diamond [35] is questioned by its observation in this work in undoped, highly pure substrates. Furthermore, no explicit observation in irradiated or ion-implanted diamond is present in the scientific literature to support its attribution to an intrinsic radiation-induced defect. The peak could thus be interpreted either as a Mg-vacancy complex different from the MgV center, in analogy with the 593.5 nm line in Sn-implanted diamond [15], or, less likely, to the predicted ground-state splitting of the MgV defect [26]. In the latter case, however, the observed splitting (~52 meV) would greatly exceed the expected value of 22 meV.

A further temperature dependent study on PL spectra (5-300 K range, 522 nm excitation, **Fig. 3b**) from the same MgV ensemble offered two additional insights in the optical activity of such a lattice complex.
Firstly, the MgV ZPL spectral shift exhibited a wavelength increase at decreasing temperatures, as shown in **Fig. 3c**. The shift was quantifiable in ~1 nm (558.4 nm) at 5 K with respect to the 557.6 nm emission at room temperature. This observation, to the best of the authors' knowledge, is unprecedented for a solid state color center, where the lattice parameter contraction at decreasing temperatures typically corresponds to a strengthening of the chemical bonds, and thus an increase in the photon energy [36-39].
This observation can be understood under the assumption that the MgV ZPL is the convolution (as indicated by the rather large ~3 nm peak FWHM) of an emission doublet. This interpretation would be in agreement with the *ab-initio* model predicting that the doublet originates from a ground state splitting [26] and that it exhibits decreasing energy separation at increasing compressive strain, and thus at decreasing temperature. The blue-shift then would be justified by an increased occupation probability of the $^4E_u$ state. High resolution spectroscopy at cryogenic temperatures will be needed to assess this hypothesis.
Secondly, the low temperature emission becomes dominated by a new set of spectral features, namely an intense peak at 608.3 nm (2.038 eV, "L1" in the following), exhibiting an intensity comparable with that of the MgV ZPL, accompanied by a set of weak bands at 619.5 nm (2.001 eV, L2), 625.3 nm (1.983 eV, L3) and 652.3 nm (1.901 eV) which can be tentatively interpreted as L1 phonon replicas.
The L1 line energy is compatible with its attribution to the $^2A_{1u} \rightarrow {}^2E_g$ transition, predicted at ~2 eV in Ref. [26]. The lack of observation of this line at room temperature is compatible with the fact that it is a weakly-allowed transition. The ~185 meV energy difference between the $^2A_{1u}$ and the $^2E_u^{(2)}$ states (i.e., the excited states of the ZPL and L1 transitions, for which the same final state $^2E_g$ is assumed [26] might be sufficient to favor the population of the latter upon room temperature phonon-assisted processes, thus quenching the L1 emission line.

*3.2 Optical excitability*

The ensemble spectral emission of the same Mg-implanted region was also investigated under different laser excitation wavelengths (**Fig. 4a-d**), namely 405 nm, 450 nm, 490.5 nm, 509.5 nm (2.2mW) and 522 nm using a fiber-coupled single-photon-sensitive confocal microscope (details on optical filtering and position of the Raman scattering lines are given **Table A1** in the Methods Section).
All the above-mentioned excitation wavelengths highlighted the occurrence of the ZPL and P2-P3 emission lines, thus confirming their attribution to photoluminescence features in Mg-implanted diamond rather than Raman-related features of the implanted host material. Differently with respect to the other laser sources, for which a 505 nm long pass filter was used, the excitation spectrum under 522 nm excitation (Fig. 1b) was acquired using a 567 nm long-pass dichroic mirror. In this case, the ~50% transmittance of the filter at 560 nm [40] was sufficient to observe the ZPL of the MgV center but partially suppressed its emission intensity. Moreover, in the case of 522 nm excitation, the first-order Raman scattering occurs at 561 nm wavelength and thus partially overlaps with the ZPL of the MgV center under the given resolution (~4 nm) of the spectrometry energy, suggesting a progressive photoionization processes, inducing a conversion between the charge state configurations available within the energy gap of diamond [26].

This hypothesis is further supported by the inability to detect single-photon emission under <510 nm excitation in regions where an abundance of individual centers were identified at longer wavelengths (see **Sect. 3.3** for 522 nm analysis of single-photon emission).

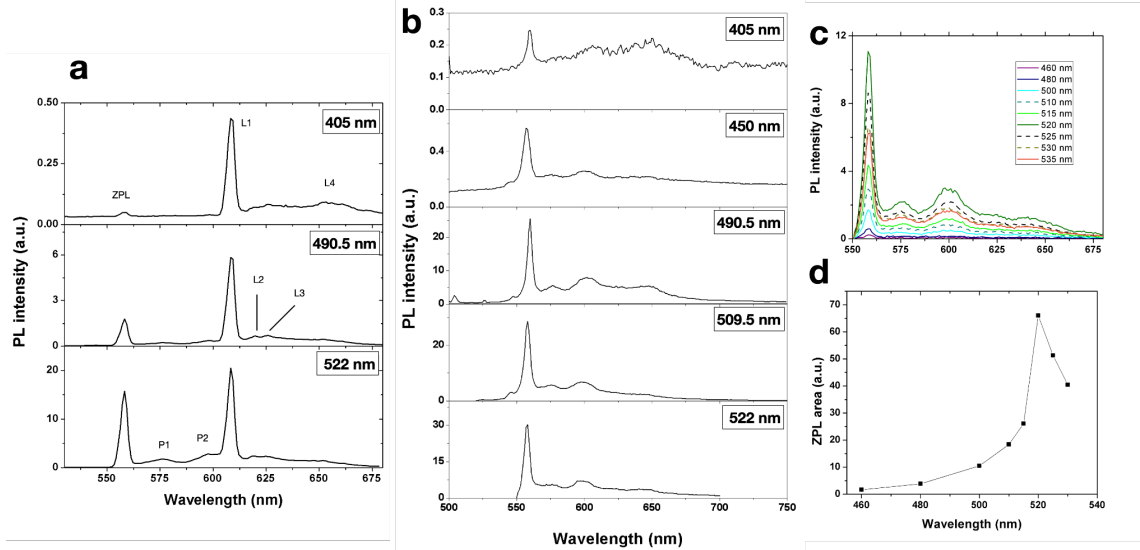

**Figure 4:** Ensemble spectra acquired: **a)** at 5 K temperature under: 405 nm (3.6 mW optical power), 490.5 nm (4.0 mW), 522 nm (3.6 mW) excitation wavelengths. **b)** at room temperature under **b)** 405 nm (8mW), 450 nm (2.2 mW), 490.5 nm (2.3 mW), 509.5 nm (2.2 mW) and 522 nm excitation wavelengths (0.3 mW). The spectra are normalized to the optical excitation power. **c)** background subtracted spectra acquired under 50 µW pulsed laser excitation (80 MHz repetition rate). **d)** Excitation wavelength dependence of the 556.7 nm ZPL peak area extracted from the spectra in **Figure 2c**.

A further investigation of the dependence of the Mg-related PL on the excitation energy was performed using a dedicated supercontinuum laser source filtered in order to have an emission bandwidth of 10 nm with central wavelength in the range 460-535 nm (**Appendix A1** for further details). The different ensemble spectra in **Fig. 4c** were acquired with 50 µW fixed optical power. (the data are background subtracted, the background was measured acquiring spectra in a pristine region of the sample to remove the contribution originating from Raman scattering). The area subtended by the ZPL peaks, reported in **Fig. 4d,** confirms that the maximum excitability of the MgV emission is at 522 nm, as suggested also by **Fig. 4a-b**. This feature indicates that the chosen wavelength is an effective tool to provide Raman-resonant excitation of the MgV center, thus enabling, in perspective, an efficient resource to maximize its emission intensity and optically address and coherently control individual emitters [41] using readily available and cost-effective laser diodes.

*3.3. Single-photon emission analysis*

PL measurements on individual MgV centers were performed under 522 nm excitation at the outer edge of a region implanted at a $5 \times 10^{12}$ cm$^{-2}$ fluence. **Figure 6a** shows a 7x8 µm$^2$ map of such a region, acquired in confocal microscopy (1 mW excitation power), where the formation of individual luminescent spots ~1 µm$^{-2}$ density is clearly recognizable. A systematic analysis on such individual luminescent spots via Hanbury-Brown & Twiss interferometry enabled to identify and analyze single-emitting defects, whose room-temperature characterization was performed in terms of spectral properties, radiative lifetime of the excited state and intensity saturation emission.

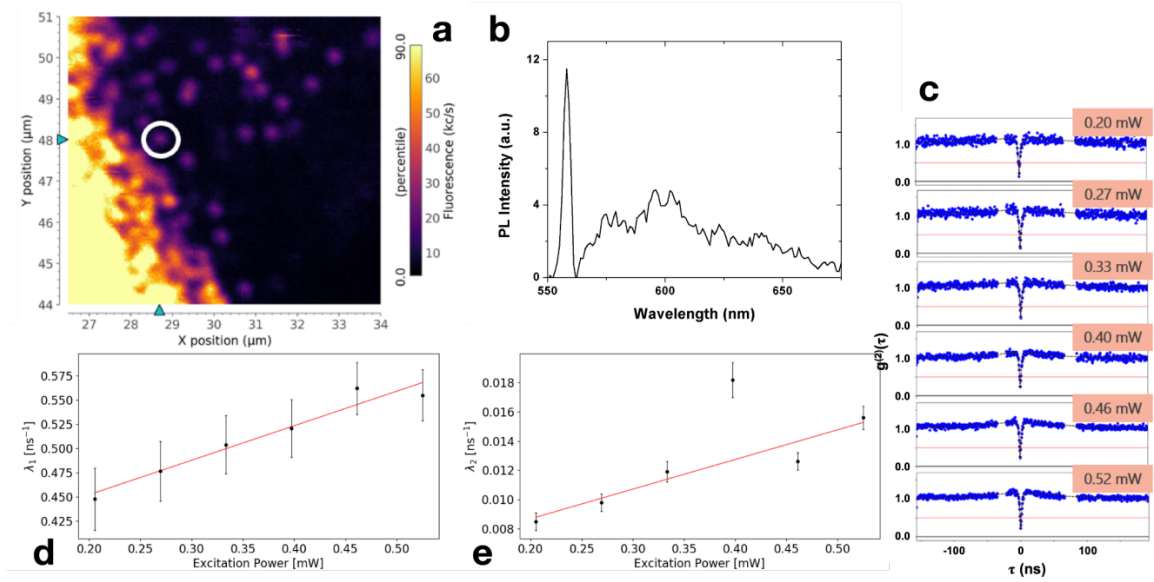

**Figure 5:** Room-temperature single-photon emission from MgV centers: **a)** Confocal microscopy map acquired under 522 nm laser excitation (1.0 mW) at the outer edge of a region implanted with Mg$^+$ ions at $5 \times 10^9$ cm$^{-2}$ fluence. **b)** Background-subtracted PL spectrum acquired from the individual emitter circled in white in Fig. 3a. **c)** Second-order auto-correlation function measurements of the same emitter under different optical excitation powers (0.2-0.52 mW range). **d)** Dependence of the $\lambda_1$ emission parameters on the optical excitation power. **e)** Dependence of the $\lambda_2$ emission fitting parameter on the optical excitation power.

As an example, the individual emitter highlighted in **Figure 5a** by a white circle has the emission spectrum shown in **Fig. 5b** upon subtraction of the background emission acquired in the same experimental conditions from an unimplanted region of the sample) to discriminate the ZPL from the diamond Raman signal, i.e. 561.0 nm and ~600 nm for first- and second-order scattering.

The measurement of the second-order autocorrelation function $g^{(2)}(\tau)$, presented in **Fig. 5c**, was performed under different optical power excitations -0.20-0.52 mW range- highlighting a strong bunching at a null delay time. This behavior, which indicates the occurence of non-classical emission, resulted in a background-subtracted value of $g^{(2)}(\tau=0)$ well below 0.5 (0.15±0.10 for 0.27 mW power), which is regarded as the threshold for single-photon emitter discrimination. A bunching effect also became visible at increasing excitation powers, suggesting the presence of a shelving state involved in the emission dynamics of the center. The background-subtracted $g^{(2)}(\tau)$ curves were thus fitted (black dashed lines in **Fig. 5b**) according to the $g^{(2)}(\tau)$ model corresponding to a three-level system

$$g^{(2)}(\tau) = 1 - (1 + a)\cdot\exp(-|\tau|\cdot\lambda_1) + a\cdot\exp(-|\tau|\cdot\lambda_2) \qquad (1)$$

where $\lambda_1$ and $\lambda_2$ are reciprocal of the characteristic times associated with the de-excitation of the excited state and the shelving state, respectively [42]. The radiative lifetime ($1/\lambda_1$) of the center was finally estimated as (2.7±0.3) ns (**Fig. 5d**) [43]. A statistical analysis based on 15 individual emitters acquired from the same region (**Fig. 6a**) revealed a relatively small dispersion of the data in the 2-3 ns range, with a weighted average lifetime of (2.4±0.2) ns, in good agreement with the preliminary data reported in Ref. [25]. The fitting function in eq. (1) enabled also to investigate the dependence of the $\lambda_2$ parameter on the power of the optical excitation. **Fig. 5e** shows the data acquired for the same emitter considered in **Fig. 5c-d**. Notably, a clear increasing trend was observed, and the behavior was consistent for all the 15 emitters analyzed in this work. This behavior, rarely reported in literature, suggests the existence of a fourth energy level interacting with the emitting system [43,44], in qualitative agreement with the findings of the theoretical model describing the opto-physical properties of the MgV center. Indeed, in Ref. [26] the photostability of the defect is attributed to the continuous induction of photoionization processes upon optical excitation. The emission intensity of individual MgV centers was also studied as a function of the optical power of the laser excitation. **Fig. 6b** displays the large variability (up to 50% of the distribution average) of the background-subtracted emission rate in saturation regime for a subset of the emitters analyzed in **Fig. 6a**. The data points were fitted according to the saturation model [11,45,46]

$$I(P) = I_{sat} \cdot P / (P + P_{sat}) \qquad (2)$$

in which I is the emission rate and P is laser excitation power. The saturation intensity cannot be regarded as a reliable estimate of the brightness of the MgV center, since a significant part of its ZPL emission is filtered by the 567 nm long-pass dichroic mirror adopted for the confocal microscope implementation. Even without considering the 50% transmittance of the latter at 560 nm, the PL count rate at saturation ranges between (0.44±0.03) Mcps and (1.46±0.17) Mcps (blue and red data in Fig. 7b, respectively). Conversely, the saturation power under 522 nm excitation was estimated in the range (0.87±0.09)–(2.7±0.4) mW. In general, a qualitative trend correlating increasing emission intensity and saturation power could be observed.

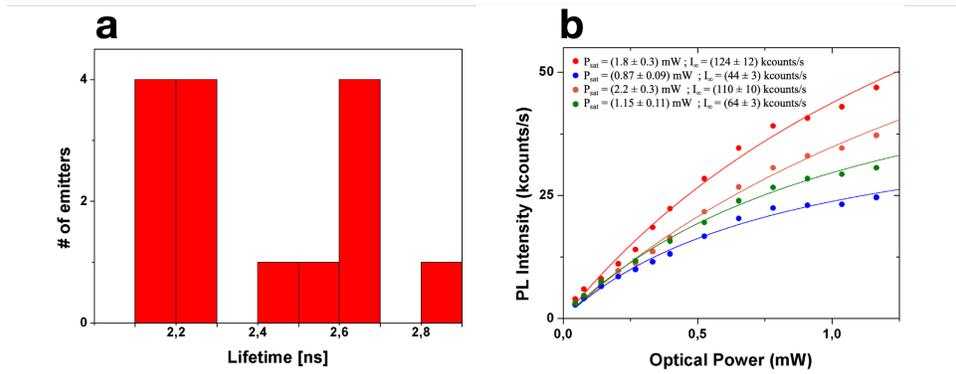

**Figure 6: a)** Distribution of the excited state lifetime of 15 individual MgV centers at room temperature. **b)** Background-subtracted emission rate as a function of the optical excitation power (522 nm wavelength). The colored curves correspond to an exemplary subset of the emitters studied in Fig. 4a. The legend indicates the fitting parameters obtained according to Eq. (2).

The single-photon emission was also investigated at cryogenic temperatures. **Fig. 7** shows the typical results obtained for individual color centers at 7 K under 522 nm, excitation. The identification of single MgV emitters was validated by the measurement of the PL spectral (**Fig. 7a**), the evaluation of the $g^{(2)}(0)$ parameter (**Fig. 7b**) (nominally, 0.05±0.01 upon background removal [17]) and the excited state lifetime evaluation (**Fig. 7c**) (2.8±1.1 ns). We underline that these results are compatible with what observed in the measurement we carried on at room temperature. Interestingly, the intense L1 emission feature observed from ensemble measurements at cryogenic temperatures could not be identified at the single-photon emitter level. Conversely, dedicated studies at varying excitation wavelengths did not succeed at isolating individual color centers, thus confirming the ensemble analysis presented in **Fig. 4**.

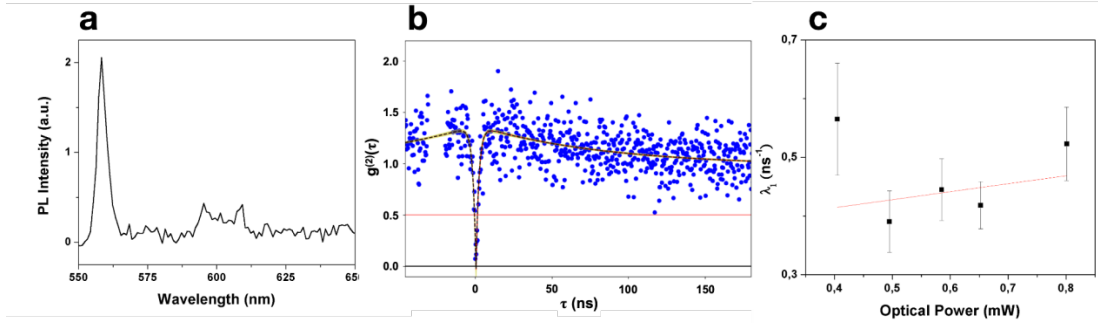

**Figure 7:** Investigation of single-photon emission from MgV centers at 7 K temperature: **a)** Background-subtracted PL spectrum acquired from an individual emitter under 522 nm optical excitation (4.6 mW); **b)** second-order autocorrelation function acquired from the same emitter under 0.7 mW excitation power; **c)** dependence of the $\lambda_1$ emission parameters on the optical excitation power. The excited lifetime for the emitter was estimated as (2.8±1.1) ns according to eq. (1).

## Conclusions

This work offers a first systematic experimental analysis of the structural and photo-physical properties of MgV centers fabricated upon ion implantation, following the first report on their optical activity [25] and the development of a first theoretical model [26]. The emission channeling analysis of $^{27}$Mg-implanted diamond revealed a high fraction of Mg (30-42%) in sites that are compatible with the split-vacancy configuration proposed as the structure of the relevant optically-active center. This indicates the very efficient structural formation of MgV, which appears comparable to that of the SnV defect [31]. The spectral features of MgV centers at both room-temperature and cryogenic temperatures also confirmed the signature of ZPL at 557.6 nm attributed to the $^2E_u^{(2)} \rightarrow {}^2E_g$ optical transition. The temperature dependence of this emission line showed a peculiar, counter-intuitive blue-shift, which was interpreted in terms of an increased occupation probability of the $^4E_u$ state.

The analysis also showed previously unexplored emission properties, including the occurrence of a previously unreported line at 608.3 nm at cryogenic temperatures, compatible with the $^2A_{1u} \rightarrow ^2E_g$ transition predicted in Ref. [26], and the persistence of the 544 nm emission under all the considered excitation wavelengths in undoped substrates (in disagreement with the results in [35]).

Finally, the single-photon emitter analysis of MgV centers suggests an appealing excited state radiative lifetime (2.7±0.3) ns, paired with a bright emission rate in reaching, in the saturation regime, up to (1.46±0.17) Mcps under (2.7±0.4) mW Raman-resonant excitation power, despite a large variability in the data distribution. Our results also suggest that the emission may be properly described by a four-level system model.

The intense emission, the high Debye-Waller factor and the enticing spin properties predicted for the center's ground state [26] offer a relevant perspective of practical applications of the MgV center in quantum technologies, which is further enhanced by its high efficiency formation exploiting ion implantation.

**Acknowledgements**

This work was supported by the following projects: 'Intelligent fabrication of QUANTum devices in DIAmond by Laser and Ion Irradiation' (QuantDia) project funded by the Italian Ministry for Instruction, University and Research within the 'FISR 2019' program; 'Training on LASer fabrication and ION implantation of DEFects as quantum emitters' (LasIonDef) project funded by the European Research Council under the 'Marie Skłodowska-Curie Innovative Training Networks' program; experiments ROUGE and QUANTEP, funded by the 5th National Commission of the Italian National Institute for Nuclear Physics (INFN); Project "Piemonte Quantum Enabling Technologies" (PiQuET), funded by the Piemonte Region within the "Infra-P" scheme (POR-FESR 2014-2020 program of the European Union); "Departments of Excellence" (L. 232/2016), funded by the Italian Ministry of Education, University and Research (MIUR); "Ex post funding of research - 2021" of the University of Torino funded by the "Compagnia di San Paolo"; The projects 20IND05 (QADeT) and 20FUN05 (SEQUME) leading to this publication have received funding from the EMPIR programme co-financed by the Participating States and from the European Union's Horizon 2020 research and innovation programme. J.F. gratefully acknowledge the EU RADIATE Project (grant ID 824096; proposal 20002351-ST) for granting transnational access to the IMBL laboratory at KU Leuven. A.C., G.M., J.M., S.M.T., R.V., A.V., L.M.C.P. acknowledge support from the KU Leuven and the Research Foundation – Flanders (FWO, Belgium).

We appreciate the support of the ISOLDE Collaboration and technical teams. This work was funded by the Portuguese Foundation for Science and Technology (Fundação para a Ciência e a Tecnologia FCT, CERN/FIS-TEC/0003/2019). The EU Horizon 2020 Framework supported ISOLDE beam times through Grant Agreement No. 654002 (ENSAR2).

The authors would like to express their gratitude to Optoprim Srl and nLight for their collaboration and technical support.

**Methods**

The experiments were performed using a micro-Raman optical setup equipped with a Horiba JobinYvon HR800 spectrometer with 2 different gratings (1800 and 600 grooves/mm) and a cooled CCD detector, Two polarized laser are mounted in the setup consisting in a 532 nm wavelength with an optical power of 250 mW and 633 nm wavelength with an optical power of 20 mW. A set of objectives (10x, 20x, 50x, 100x) are mounted on a OlympusBx41 microscope which is arranged to operate in both reflection and transmission. Other excitation wavelengths were explored by means of two custom fiber-coupled single-photon sensitive confocal microscopes. In such systems a multimode optical fiber (core diameter ø=50 μm) was used both as the pinhole of the confocal microscope and as the outcoupling medium for luminescence detection and analysis (**Fig. A1.a**). Room temperature measurements were performed using a 100× air objective (0.9 NA), the sample was scanned over a 100×100×100 μm$^3$ closed-loop piezoelectric nanopositioner. Cryogenic measurements were performed by interfacing the aforementioned custom confocal microscope to a Montana S100 cryostation (**Fig. A1.b**), equipped with a vacuum-compatible long-distance air objective (100×, 0.85 N.A.). The sample was mounted on a three-axes open-loop nanopositioner. Luminescence maps and photon count rate were monitored by means of the QUDI open source software [47].

Second-order autocorrelation function measurements were performed by a Hanbury-Brown & Twiss interferometer implemented by connecting a fiber-fused balanced beam-splitter to two independent commercial silicon single-photon avalanche detectors (SPAD, 250 cps dark counts). Time correlated single-photon counting (and the relative concidence measurements) were performed exploiting a ID800 ID Quantique time tagger.

Spectral measurements were performed by connecting the optical output of the confocal microscope to a Horiba iHR320 monochromator, whose output port was fiber-coupled to a SPAD (spectral resolution: ~4 nm [24]). Measurements at varying laser wavelength were performed on the room-temperature PL setup by using both a discrete set of laser diodes and a supercontinuum laser (80 MHz NKT SuperK Fianum) with tunable emission wavelength (<10 nm bandwidth, **Fig. 4c**).

| Excitation source | Confocal microscope | Spectral filtering | Raman line |
|---|---|---|---|
| 405 nm | Single-photon sensitive | 505-700 nm spectral bandwith | 428.1 nm |
| 450 nm | Single-photon sensitive | 505 nm long pass filter | 478.7 nm |
| 490.5 nm | Single-photon sensitive | 505 nm long pass filter | 524.8 nm |
| 509,5 nm | Single-photon sensitive | 525 nm long pass filter | 546.5 nm |
| 522 nm | Single-photon sensitive | 550 nm long pass filter | 561.0 nm |
| 532 nm | Micro-Raman | Notch filter at 532 nm | 572.8 nm |

**Table A1.** List of the laser excitation wavelengths adopted for the characterization of MgV centers, including details on the spectral filtering and the spectral position of the corresponding first-order Raman line.